\RequirePackage[2020-02-02]{latexrelease} 
\documentclass[twocolumn]{aastex62}
\graphicspath{{./}{figures/}}

\submitjournal{ApJL}

\shorttitle{JWST on SMG stellar morphology}
\shortauthors{Chen et al.}

\begin{document}

\title{JWST sneaks a peek at the stellar morphology of $z\sim2$ submillimeter galaxies: Bulge formation at cosmic noon}

\correspondingauthor{Chian-Chou Chen (TC)}
\email{ccchen@asiaa.sinica.edu.tw}

\author[0000-0002-3805-0789]{Chian-Chou Chen}
\affil{Academia Sinica Institute of Astronomy and Astrophysics (ASIAA), No. 1, Section 4, Roosevelt Road, Taipei 10617, Taiwan}

\author{Zhen-Kai Gao}
\affiliation{Academia Sinica Institute of Astronomy and Astrophysics (ASIAA), No. 1, Section 4, Roosevelt Road, Taipei 10617, Taiwan}
\affiliation{Graduate Institute of Astronomy, National Central University, 300 Zhongda Road, Zhongli, Taoyuan 32001, Taiwan}

\author{Qi-Ning Hsu}
\affiliation{Graduate Institute of Astrophysics, National Taiwan University, Taipei 10617, Taiwan}
\affiliation{Academia Sinica Institute of Astronomy and Astrophysics (ASIAA), No. 1, Section 4, Roosevelt Road, Taipei 10617, Taiwan}

\author{Cheng-Lin Liao}
\affiliation{Graduate Institute of Astrophysics, National Taiwan University, Taipei 10617, Taiwan}
\affiliation{Academia Sinica Institute of Astronomy and Astrophysics (ASIAA), No. 1, Section 4, Roosevelt Road, Taipei 10617, Taiwan}

\author{Yu-Han Ling}
\affiliation{Academia Sinica Institute of Astronomy and Astrophysics (ASIAA), No. 1, Section 4, Roosevelt Road, Taipei 10617, Taiwan}

\author{Ching-Min Lo}
\affiliation{Academia Sinica Institute of Astronomy and Astrophysics (ASIAA), No. 1, Section 4, Roosevelt Road, Taipei 10617, Taiwan}

\author[0000-0003-2588-1265]{Wei-Hao Wang}
\affiliation{Academia Sinica Institute of Astronomy and Astrophysics (ASIAA), No. 1, Section 4, Roosevelt Road, Taipei 10617, Taiwan}

\author{Yu-Jan Wang}
\affiliation{Academia Sinica Institute of Astronomy and Astrophysics (ASIAA), No. 1, Section 4, Roosevelt Road, Taipei 10617, Taiwan}



\begin{abstract}

We report morphological analyses of seven submillimeter galaxies (SMGs) at $z\sim2$ using the JWST NIRCam images taken as part of the public CEERS and PRIMER surveys. Through two-dimensional surface brightness profile fittings we find evidence of bulges in all the sample SMGs, in particular at F444W filter, suggesting an ubiquitous presence of stellar bulges. The median size of these bulges at F444W is found to be 0.7$\pm$1.0\,kpc and its median Sersic index is 0.7$\pm$0.9. Structures akin to spiral arms and bars are also identified, although their asymmetric shapes, tidal features, as well as evidence of nearby galaxies at consistent redshifts as those of corresponding SMGs suggest that these SMGs are undergoing dynamical interactions, likely responsible for the triggering of their star-forming activities. Via the curve-of-growth analyses we deduce half-light radii for the NIRCam wavebands, finding that sizes are significantly smaller at longer wavelengths in all cases, in particular that the median size ratio between F444W and F150W is $0.6\pm0.1$. However, we also find that F444W sizes, roughly corresponding to rest-frame $H$-band, are not smaller than those of submillimeter continuum as measured by ALMA, contrasting recent predictions from theoretical models. Our results suggest that while stellar bulges are undergoing an active formation phase in SMGs at $z\sim2$, the total stellar masses of SMGs are still dominated by their disks, not bulges. 


\end{abstract}

\keywords{High-redshift galaxies --- Galaxy structure --- Galaxy interactions --- Galaxy evolution --- Ultraluminous infrared galaxies}


\section{Introduction}\label{sec:intro}
The epoch of $z\sim1-3$, so called the cosmic noon, represents a key phase of rapid stellar mass assembly for massive galaxies, and this active formation phase is now known to mostly take place in galaxies that are infrared luminous \citep{Madau:2014aa}. These dusty infrared luminous star-forming galaxies are often detected via submillimeter observations \citep{Smail:1997p6820,Barger:1998p13566,Hughes:1998p9666,Blain:2002p8120,Casey:2014aa}, and these submillimeter galaxies, or the SMGs, are massive systems at cosmic noon (e.g., \citealt{Chapman:2005p5778,Danielson:2017aa,Chen:2022aa}) that are believed to intimately link to compact quiescent galaxies at similar redshifts and the formation of massive local ellipticals \citep{Lilly:1999lr,Toft:2014aa,Dudzeviciute:2020aa}. Given their important role in galaxy evolution, exactly how SMGs build up their stellar mass is one key question to galaxy formation models.

\begin{figure*}[ht!]
	\begin{center}
		\leavevmode
		\includegraphics[scale=0.35]{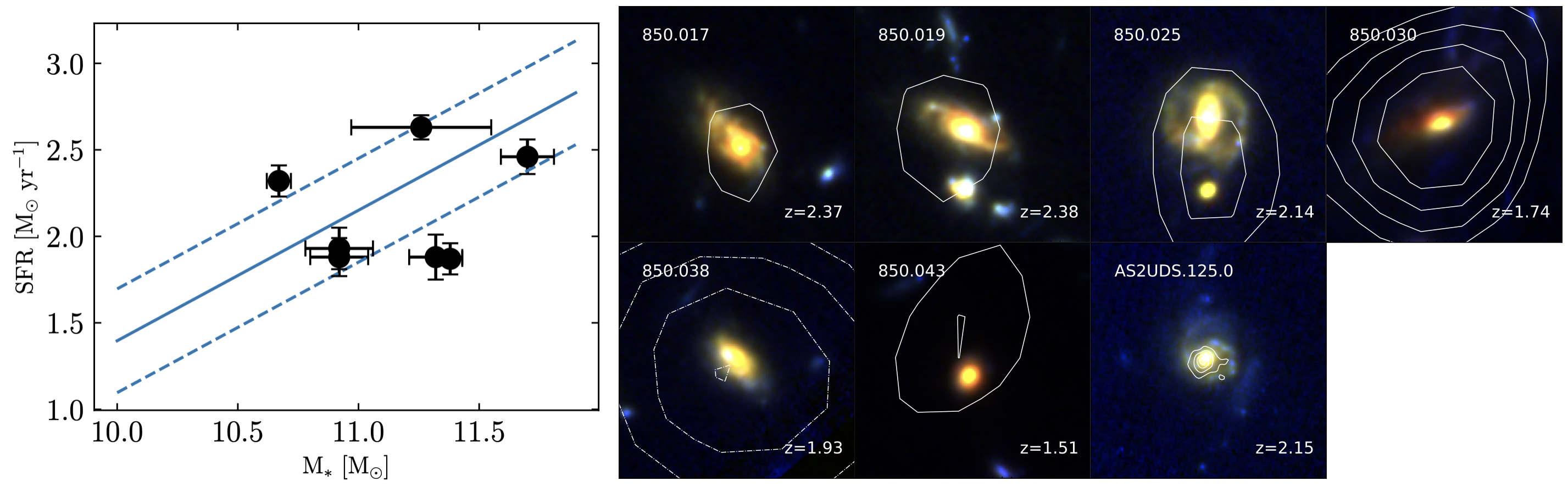}
		\caption{{\it Left:} Stellar mass - SFR main sequence at $z=2$ based on \citep{Speagle:2014aa}. Dashed lines show the range of $\pm$0.3\,dex scatter. The locations of the sample SMGs are marked as black data points. Our sample SMGs lie on the main sequence at $z\sim2$. {\it Right:} Thumbnail images of our sample SMGs with their IDs shown at the top-left corner and their redshits at bottom-right. The RGB images with 40\,kpc on a side are made with NIRcam F115W, F277W, and F444W filters, except for 850.019 where the blue image is made with the F150W filter. The images are made with a linear stretch between 1 and 99.9 percent. The larger solid contours show the VLA 1.4\,GHz emissions with [3,4,5,6]$\times$$\sigma$ levels, the dashed contours show the MIPS 24\,$\mu$m emissions with [10,13,16,19]$\times$$\sigma$ levels, and finally the smaller solid contours show the ALMA 870\,$\mu$m continuum emissions with [5,10,15,20]$\times$$\sigma$ levels. The OIR counterparts of the sample SMGs are robustly identified by VLA, MIPS, or ALMA. 
		}
		\label{fig:fig1}
	\end{center}
\end{figure*}

Recent observational studies have shown almost ubiquitously that the sizes of submillimeter continuum of SMGs are compact, about 1-2\,kpc \citep{Simpson:2015aa,Ikarashi:2015aa,Hodge:2016aa,Spilker:2016aa,Fujimoto:2018aa,Gullberg:2019aa}. The submillimeter sizes are also found to be smaller than those deduced from optical and near-infrared continuum images \citep{Chen:2015aa,Fujimoto:2018aa,Lang:2019aa} or emission lines \citep{Calistro-Rivera:2018aa,Chen:2020aa}, typically by a factor of 2-3. These evidence suggest that SMGs are actively forming stars in the central regions and the stellar bulges have been quickly built up. Bulge formation, or similarly the inside-out growth scenario, has been suggested from various studies for massive galaxies at cosmic noon (e.g., \citealt{Nelson:2016ab,Tadaki:2020ab}), and has been predicted also by recent theoretical models (e.g., \citealt{Cochrane:2019aa,Popping:2022aa}. However, due to heavy dust obscuration, observational evidence of the presence of stellar bulges in SMGs has been an indirect one, including those inferred from the observed colors \citep{Lang:2019aa}. Heavy dust obscuration has also impacted galaxy size measurements in general, and recent studies using color information has suggested that the true stellar sizes are significantly smaller than what have been inferred from {\it HST} $H$-band studies \citep{Suess:2019aa,Miller:2022aa}.

On the other hand, heavy dust obscuration has also hampered our ability to determine the triggering mechanism of star formation in SMGs using {\it HST} images, where often times it is not possible to distinguish irregular disks and mergers \citep{Swinbank:2010aa,Mortlock:2013aa,Chang:2018aa}. However the dominant triggering mechanism of star formation for SMGs has been a key factor that differentiates various theoretical models \citep{Baugh:2005p14519,Dave:2010kx,Hayward:2013qy,Cowley:2015aa,Lagos:2020aa}, making a definite determination on this issue would make a major step forward in modeling SMGs, as well as the formation of massive galaxies in general. 

With the advent of JWST, its sensitive high resolution observations at mid-infrared allows us to finally be able to directly image rest-frame near-infrared morphologies for galaxies at cosmic noon, revealing stellar components that are previously hidden in images at shorter wavelengths. Indeed, early results from JWST have found that the mid-infrared sizes are smaller than those of previous {\it HST} $H$-band observations \citep{Cheng:2022aa,Suess:2022aa}, suggesting significant amount of missing stellar emissions in the central regions in previous measurements. 

In this Letter we aim to study stellar distributions of SMGs at $z\sim2$ using JWST images. We describe sample selection and data in \autoref{sec:obs}. We present our analyses and results in \autoref{sec:results}, and finally discussion and summary are given in \autoref{sec:summary}. Throughout this paper
we assume the Planck cosmology: $H_0$ = 67.7~km~s$^{-1}$~Mpc$^{-1}$, $\Omega_{\rm M}$ = 0.31, and $\Omega_{\Lambda}$ = 0.69 \citep{PlanckCollaboration:2020aa}.

\begin{table*}
	\caption{Properties of the sample SMGs and their measured sizes}
	\label{tab:table1}   
	\begin{center}
		\begin{tabular}{rlrllllllll}
			\hline
			ID$^a$ & $S_{\rm 850/870}^b$ & $z^c$ & log($M_{\ast}$)$^c$ & log($L_{\rm IR}$)$^c$ & $R_{\text{e,f115w}}^d$ & $R_{\text{e,f150w}}^d$ & $R_{\text{e,f200w}}^d$ & $R_{\text{e,f277w}}^d$ & $R_{\text{e,f356w}}^d$ & $R_{\text{e,f444w}}^d$\\
			& [mJy] &  & [M$_{\odot}$] & [L$_{\odot}$] & [kpc] & [kpc] & [kpc] & [kpc] & [kpc] & [kpc]\\
			\hline
			850.017 & 2.7$\pm$0.4 & 2.37 & 11.3$\pm$0.3 & 12.6$\pm$0.1 & 8.1$^{+0.6}_{-0.6}$ & 7.7$^{+0.4}_{-0.4}$ & 6.1$^{+0.2}_{-0.2}$ & 5.3$^{+0.3}_{-0.3}$ & 4.6$^{+0.2}_{-0.2}$ & 4.1$^{+0.2}_{-0.2}$\\
			850.019 & 3.4$\pm$0.5 & 2.38 & 10.7$\pm$0.1 & 12.3$\pm$0.1 & \ldots{} & 6.5$^{+0.4}_{-0.4}$ & 6.0$^{+0.3}_{-0.3}$ & 5.2$^{+0.4}_{-0.4}$ & 4.9$^{+0.3}_{-0.3}$ & 4.7$^{+0.2}_{-0.2}$\\
			850.025 & 2.7$\pm$0.3 & 2.14 & 11.3$\pm$0.1 & 11.9$\pm$0.1 & 5.9$^{+0.7}_{-0.7}$ & 5.3$^{+0.3}_{-0.3}$ & 5.0$^{+0.2}_{-0.2}$ & 4.7$^{+0.2}_{-0.3}$ & 4.4$^{+0.2}_{-0.2}$ & 4.3$^{+0.2}_{-0.2}$\\
			850.030 & 2.1$\pm$0.4 & 1.74 & 11.4$\pm$0.1 & 11.9$\pm$0.1 & 11.3$^{+0.7}_{-0.7}$ & 10.3$^{+0.7}_{-0.7}$ & 7.8$^{+0.8}_{-0.8}$ & 6.7$^{+0.8}_{-0.8}$ & 6.1$^{+0.8}_{-0.7}$ & 5.8$^{+0.7}_{-0.7}$\\
			850.038 & 2.1$\pm$0.4 & 1.93 & 10.9$\pm$0.1 & 11.9$\pm$0.1 & \ldots{} & 5.0$^{+0.3}_{-0.3}$ & \ldots{} & 4.7$^{+0.3}_{-0.3}$ & 4.2$^{+0.2}_{-0.2}$ & 3.9$^{+0.2}_{-0.2}$\\
			850.043 & 1.7$\pm$0.4 & 1.51 & 10.9$\pm$0.1 & 11.9$\pm$0.1 & 3.0$^{+0.3}_{-0.3}$ & 2.6$^{+0.1}_{-0.1}$ & 2.2$^{+0.1}_{-0.1}$ & 1.8$^{+0.1}_{-0.1}$ & 1.6$^{+0.1}_{-0.1}$ & 1.6$^{+0.1}_{-0.1}$\\
			AS2UDS.125.0 & 4.6$\pm$0.5 & 2.154 & 11.7$\pm$0.1 & 12.5$\pm$0.1 & 7.0$^{+0.7}_{-0.7}$ & 7.1$^{+0.3}_{-0.3}$ & 5.2$^{+0.1}_{-0.1}$ & 5.1$^{+0.2}_{-0.2}$ & 4.6$^{+0.1}_{-0.1}$ & 4.5$^{+0.1}_{-0.1}$\\
			\hline
		\end{tabular}
		\begin{tabular}{l}
			Note: \\
			$^a$: IDs adopted from \citet{Zavala:2017aa} except for AS2UDS.125.0, which is based on \citet{Stach:2019aa}. \\
			$^b$: Deboosted 850\,$\mu$m flux densities measured from SCUBA-2 as reported by \citet{Zavala:2017aa}, except for AS2UDS.125.0, \\ 
			where its 870\,$\mu$m flux density was reported by \citet{Stach:2019aa}. \\
			$^c$: Redshifts, stellar masses, and infrared luminosities adopted from \citet{Zavala:2018aa}, except AS2UDS.125.0, for which \\
			its properties are adopted from \citet{Lang:2019aa}. The redshift of AS2UDS.125.0 is based on spectroscopic measurements. \\
			$^d$: Half-light radii ($R_{\rm e}$) are estimated from the curve-of-growth analyses (\autoref{sec:cog}).\\
			
		\end{tabular}    
	\end{center}
\end{table*}

\section{Observations and data}\label{sec:obs}

\subsection{Sample}\label{subsec:sample}
Our SMG sample is based on the SCUBA-2 cosmological survey \citep{Geach:2017aa} in the EGS and UDS fields. For the EGS field, number counts, counterpart identifications, and detailed studies of the physical properties of the submillimeter sources have been performed and presented in \citet{Zavala:2017aa,Zavala:2018aa}. We adopt their counterpart identifications and various physical properties, but we limit our selection to the sources that have both 850 and 450\,$\mu$m detection so the submillimeter detection is more robust. For the UDS field, ALMA follow-up observations on a flux-limited sample of SCUBA-2 detected submillimeter sources have been carried out and the physical properties including their precise locations have been presented in \citet{Stach:2018aa}, \citet{Stach:2019aa}, and \citet{Dudzeviciute:2020aa}.

We first select SMGs that are at redshifts between 1.5 and 2.5. The choice of this redshift range is motivated by the fact that the observed wavelengths of JWST roughly corresponds to UV for F115W and 1.6\,$\mu$m for F444W filters, allowing us to study the distributions of young and less obscured star-forming regions as well as the bulk of stellar mass. We then match these sources to the footprint of the publicly released CEERS and PRIMER data. We find one and six SMGs within the PRIMER and CEERS footprint, respectively. The properties of our seven sample SMGs are provided in \autoref{tab:table1}. As shown in \autoref{fig:fig1} they lie on the main sequence of star-forming galaxies at $z\sim2$. In \autoref{fig:fig1} we also show the thumbnail JWST images of our sample SMGs.

Overall, the median redshift of this sample is $2.1\pm0.2$, and the median 850 or 870\,$\mu$m flux density (S$_{850/870}$) is $2.7\pm0.4$\,mJy. Thus the sample is consistent with the median redshift of the general SMG population \citep{Chapman:2005p5778,Dudzeviciute:2020aa} but lies at the fainter end compared to the typical SMG samples.

\subsection{HST data}
For various purposes we make use of the publicly available CANDELS data \citep{Grogin:2011fj,Koekemoer:2011aa}, in particular the WFC3 F125W and F160W images, and the associated catalogs and estimated physical properties such as redshifts \citep{Galametz:2013aa,Stefanon:2017aa}. As a check on the CANDELS astrometry, by cross matching the CANDELS catalog to that of GAIA DR3 we confirm that their coordinates are consistent with each other to 0$\farcs$01 level. 

\subsection{JWST data}\label{subsec:jwst}
The JWST data were taken as part of the ERS program CEERS \citep{Finkelstein:2022aa}] and the public GO program PRIMER (PI: J. Dunlop). The wideband NIRCam imaging data taken using filters F115W, F150W, F200W, F277W, F356W, and F444W are obtained from the MAST archive. All the JWST data used in this paper can be found in MAST: \dataset[10.17909/f8cx-p769]{http://dx.doi.org/10.17909/f8cx-p769}. The publicly released version of the maps that we use all include stage 3 data products, which are based on images that are calibrated and resampled. We examine the images in the following and the results show that most of them are sufficient in quality for our purposes. We therefore adopt the public version and do not rerun data reduction.

We tie the astrometry of the NIRCam images to the publicly released CANDELS catalogs and find in general a pointing accuracy of $\sim$0$\farcs$1, which is consistent with what has been reported by the JWST commissioning team \citep{Rigby:2022aa}. We additionally find that the pointing offsets are significantly different between each of the two NIRCam modules, and the different offsets appear uncorrelated and it does not appear to be caused by the rotation parameters in the header. On the other hand the detectors in the same module have consistent offsets therefore no evidence of significant image distortion that would affect morphological analyses. Therefore for our analyses we make astronometry corrections to the module where our sample SMGs are located.

To validate the photometry of NIRCam imaging we run SExtractor on all the available NIRCam images and compare the photometry between NIRCam and the published catalog from CANDELS. In particular we compare the NIRCam F150W measurements with those interpolated between WFC3 F140W and F160W measurements. We also compare NIRCam F356W and F444W photometry and that of IRAC Ch1 and Ch2 in the CANDELS catalog. In all three cases, for our sample SMGs, as well as most of the galaxies on the images that can be matched, the flux density measurements agree with each other. 

We assess the depths of these NIRCam images by randomly positioning circular apertures in regions where no significant detection is obtained in SExtractor. We find that in general the 10\,$\sigma$ depths for point sources \footnote{We use 0$\farcs$08 and 0$\farcs$16 radius for short and long wavelength filters, respectively, following https://jwst-docs.stsci.edu/jwst-near-infrared-camera/nircam-performance/nircam-imaging-sensitivity.} are about 27-29 AB magnitude, consistent with expectations given their exposure times which based on the header are about 3-50\,ks. 

\section{Analyses and Results}\label{sec:results}
\subsection{Galfit}
We first employ {\sc galfit} \citep{Peng:2010aa} to analyze NIRCam images. For each image, we feed to {\sc galfit} the science maps, weight maps, Nyquist sampled point spread functions (PSFs), and source masks. For the weight maps the publicly released version are adopted but scaled in a way such that the standard deviation of the signal-to-noise ratio in blank regions is one. We find typical scaling factors of about 2-6, depending on the filters. For PSFs, the pipeline produced maps have pixel scales of 0$\farcs$03 and 0$\farcs$06 for short and long wavelength filters, respectively, which are insufficient for Nyquist sampling the PSFs in most filters. In addition, it is often not possible to find many bright and unsaturated stars within the footprint to produce high quality stacked PSFs. As a result, we adopt the following strategy that has often been used for {\it HST} images in similar conditions; We first use the {\sc WebbPSF} software \citep{Perrin:2012aa} to generate synthetic PSFs for each filter, with an oversampling factor of four. We then use {\sc galfit} to convolve the synthetic PSFs with Gaussian profiles and fit to the available nearby unsaturated stars. The resulting best-fit models are adopted as the Nyquist sampled PSFs. For source masks we adopt the aperture shapes suggested by SExtrator. The morphological parameters of these suggested apertures are also used as the first guests for the Sersic profile fitting in {\sc galfit}.

\begin{figure}[ht!]
	\begin{center}
		\leavevmode
		\includegraphics[scale=0.43]{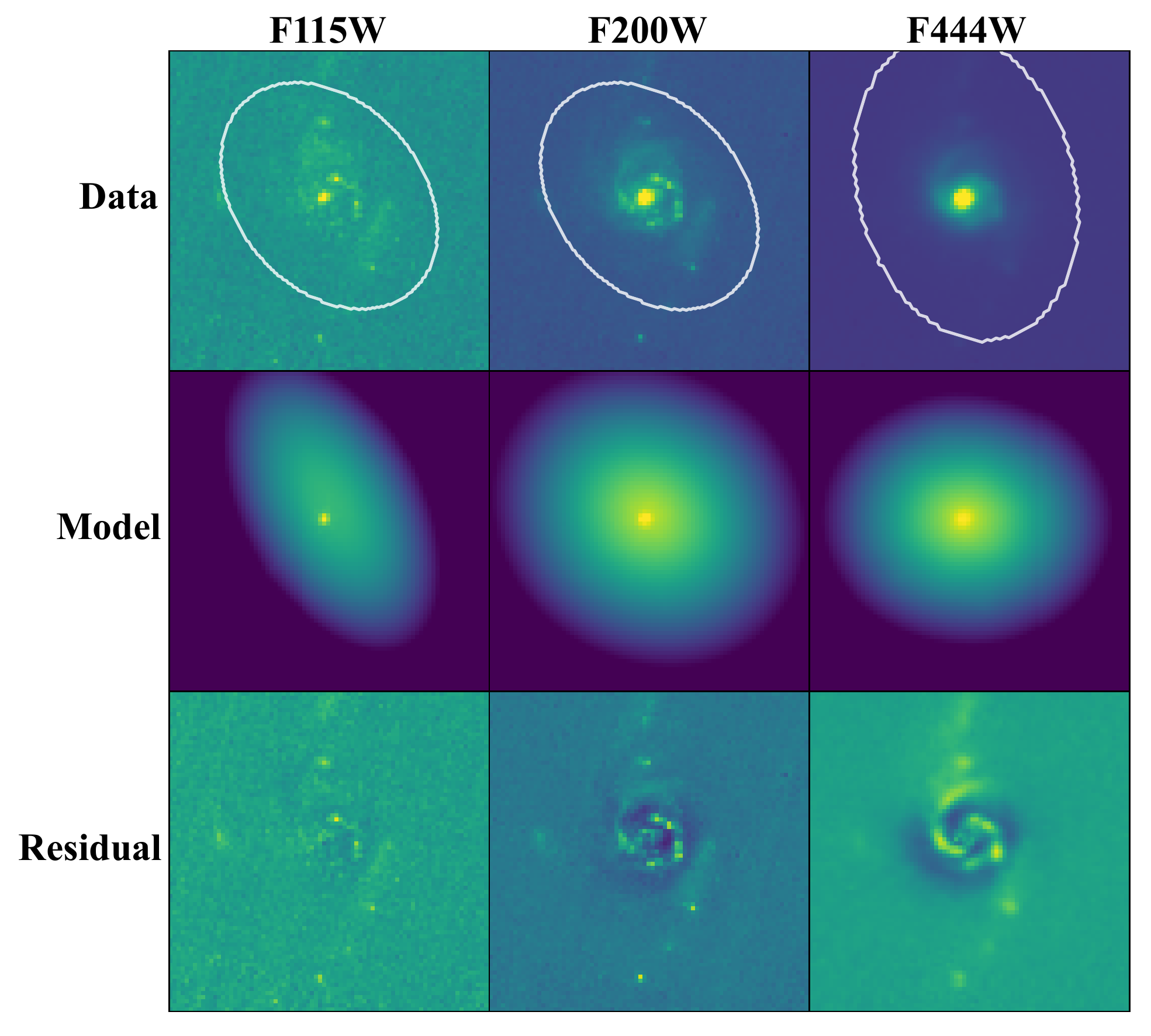}
		\caption{Top panels show the NIRCam science images with $5''$ on a side centered at the ALMA confirmed SMG AS2UDS.125.0, where the white ellipses mark the source masks within which {\sc galfit} is performed. Middle and bottom panels show the model and residual maps produced by {\sc galfit}. Example results of the three filters, F115W, F200W, and F444W are shown in each column. We find at least two components, a bulge and a disk, are required to produce adequate fitting results in almost all filters of our sample SMGs, especially for F444W in which all sources need two components (\autoref{fig:figA1}), suggesting stellar bulges are often present in SMGs.}
		\label{fig:fig2}
	\end{center}
\end{figure}

\begin{figure*}[ht!]
	\begin{center}
		\leavevmode
		\includegraphics[scale=0.9]{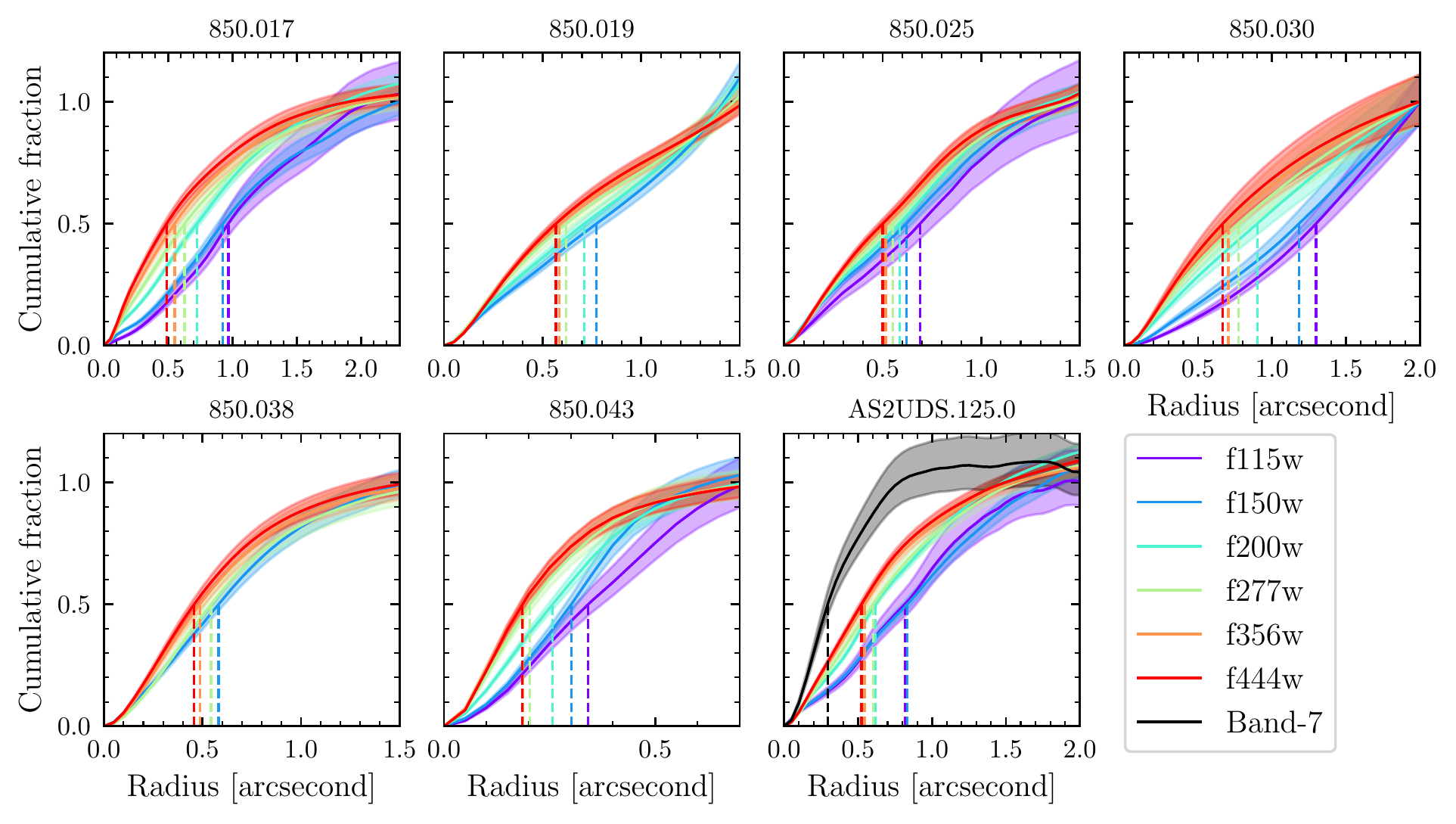}
		\caption{Cumulative fractions of integrated flux densities measured from NIRCam on all seven sample SMGs as well as those from ALMA band 7 on AS2UDS.0125.0. Vertical lines with corresponding colors mark the half-light radii. In all cases, half-light radii are smaller at longer wavelengths. It is also evident that bulges, despite presenting in many sources, do not dominate the total flux densities.
		}
		\label{fig:fig3}
	\end{center}
\end{figure*}

For each science image we always start with a single Sersic profile. We quickly find that single Sersic component often produces fits that are either not converged or large in $\chi^2$, and two components, a bulge and a disk in almost all cases, are required to significantly improve the fits. In one source, 850.025, we find that an additional bar component is also needed. Example results on AS2UDS.125.0 are shown in \autoref{fig:fig2}, and full results on F444W, which traces stellar mass distributions given the redshifts of our sample SMGs, are shown in the appendix (\autoref{fig:figA1} and \autoref{tab:tableA1}). Overall, we find the median size of the bulge in F444W filter is $0.7\pm1.0$\,kpc with a Sersic index of $0.7\pm0.9$. In addition to needing two components, we also find that in the residual maps structures mimicking spiral arms, tidal remnants, and clumps are often present. These structures suggest dynamical perturbations that could be caused by galaxy mergers or flybys. It is outside the scope of this paper to quantitatively analyze these structures but they suggest that JWST images can offer much more information than currently obtained. To avoid confusions we also leave the {\sc galfit} results in other filters to future works.

\begin{figure*}[ht!]
	\begin{center}
		\leavevmode
		\includegraphics[scale=0.6]{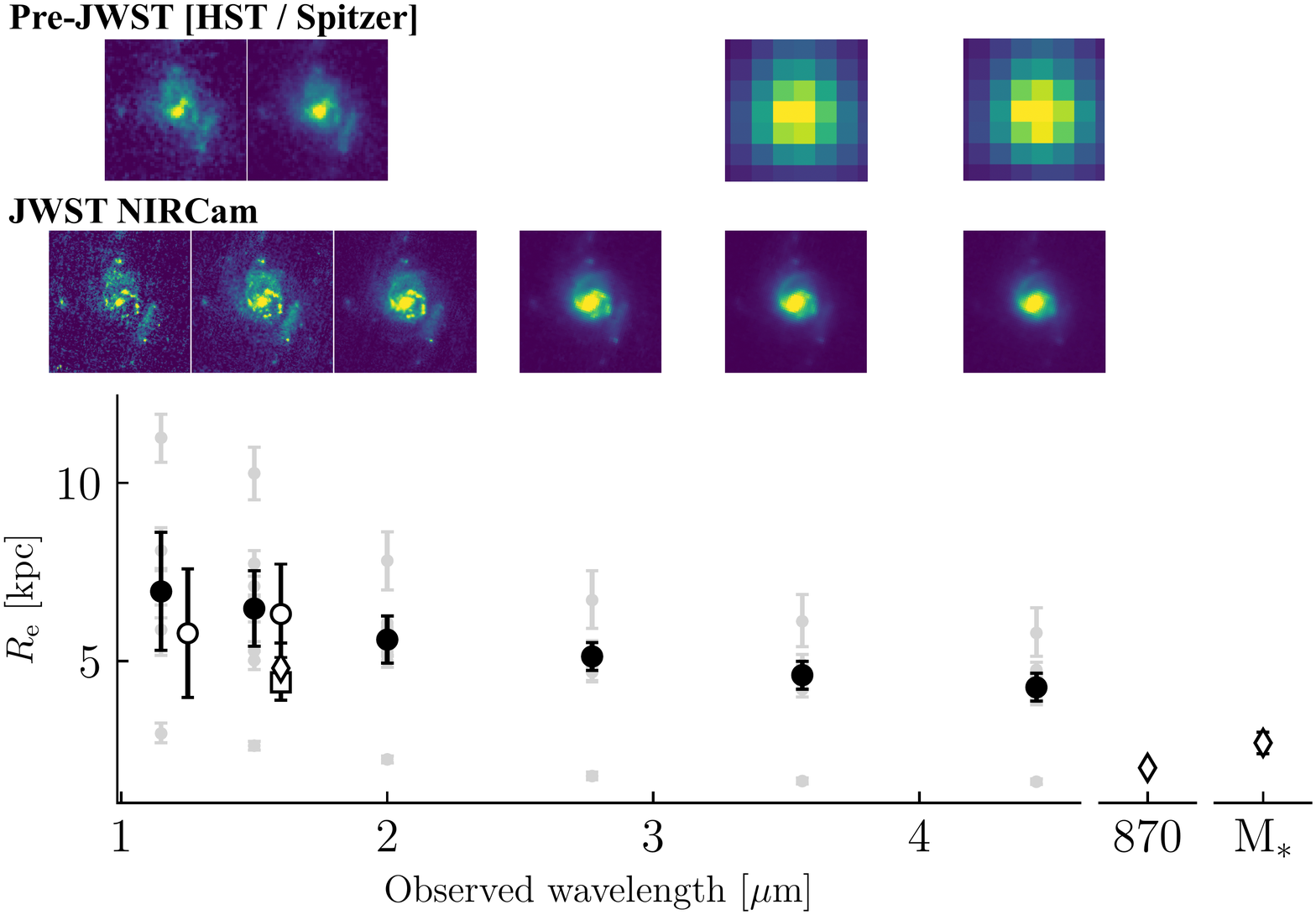}
		\caption{Half-light radii in observed wavelengths of our sample SMGs are shown in grey points. The median values of each waveband are shown in larger black points with bootstrapped uncertainties. Empty circles show size measurements of our sample SMGs based on the published {\it HST} results in $J$ and $H$-band \citep{van-der-Wel:2014aa}, empty squares show $H$-band sizes of SMGs reported by \citet{Chen:2015aa}, and empty diamonds show sizes measured by \citet{Lang:2019aa}, where the 870\,$\mu$m sizes and stellar mass sizes are also shown. On the top we show cutout images with 4$''$ on a side from {\it HST, Spitzer}, and JWST of one of our sample SMGs confirmed by ALMA, AS2UDS.125.0, and these cutouts are aligned vertically roughly to the wavelengths shown in the x-axis below. 
		}
		\label{fig:fig4}
	\end{center}
\end{figure*}

\subsection{Curve of growth}\label{sec:cog}
Since our analyses with {\sc galfit} suggest that the optical and near-infrared morphology of the sample SMGs are complicated and can not be modeled by a single Sersic profile, their overall half-light radii, or sizes, need to be estimated by other methods such as the curve-of-growth method. The apertures for the curve-of-growth analyses are constructed by adopting the positions of the peak pixel, corresponding to the bulge in most cases, as their centroids, and the shapes of the apertures are according to those suggested by SExtractor. Thus, instead of using circular apertures we adopt elliptical apertures, and the length of the major axis is what is used for the curve-of-growth.

In \autoref{fig:fig3} we show the curve-of-growth results, where the flux densities measured by each aperture sizes are divided by the total flux densities obtained from SExtractor. Images from a few filters for 850.019 and 850.038 are contaminated by uneven background at the edge of the images, leading to unreliable measurements so we do not report those results. Since the images are deep the errors of aperture photometry are small and the uncertainties of the curve-of-growth are dominated by the errors of the total flux density measurements. 

We find in all cases sources have smaller sizes at longer wavelengths, and the median half-light radii go from $0\farcs8\pm0\farcs2$ in F115W to $0\farcs5\pm0\farcs1$ in F444W. In particular, the median size ratio of $\langle R_{\rm e,F444W}/R_{\rm e,F150W}\rangle$ is found to be $0.6\pm0.1$, consistent with recent findings for general galaxy populations at the redshift range of the sample SMGs \citep{Suess:2022aa}. We provide half-light radii in each filter for each source in \autoref{tab:table1}. Note that one typical concern of using the curve-of-growth analyses to derive sizes is that it is not straightforward to take into account the broadening effect caused by the PSFs. This becomes a more serious issue when the source sizes are close to those of the PSFs. However as shown in \autoref{fig:fig3} most of our source are much larger than the PSFs, except 850.043. By forcing {\sc galfit} to perform single Sersic component fitting we determine that the true sizes of 850.043 to be smaller by 30\% at most, which occurs in F444W, and the differences are much smaller in other filters. However since we mainly discuss median values one single source does not affect our conclusions significantly.

\section{Discussion and Summary}\label{sec:summary}
In this Letter we analyze morphology and sizes of a sample of seven SMGs at $z\sim2$ using JWST NIRCam imaging. Via {\sc galfit} analyses we report detection of bulge components in all the sample SMGs, most prominently in F444W filter, which in principle traces stellar distributions at the redshift range of our sample. The detection of bulges is enabled thanks to the superior sensitivity and spatial resolution of JWST. Indeed, by examining previous {\it HST} imaging and morphological analyses on our sample SMGs by \citet{van-der-Wel:2014aa}, there already can be observed hints of bulge component, but the much lower signal-to-noise detection together with poorer spatial resolution means that Sersic profile fitting can yield adequate results with just one single component. In two cases (850.019 and 850.025) the fitting of the {\it HST} images already leans toward the bulge as suggested by the reported Sersic indices, and in two case (850.030 and AS2UDS.125.0) bad fits were reported, suggesting complicated morphology confirmed by the JWST imaging (\autoref{fig:figA1}). 

The residual images produced from {\sc galfit} are also informative. Structures mimicking spiral arms can been seen in most cases, and one SMG (850.025) requires a bar component for the fitting. These structures are consistent with those suggested by recent high-resolution ALMA observations on dust morphologies of SMGs, although those are in the central regions so on a much smaller scale \citep{Hodge:2019aa,Gullberg:2019aa}. However, the spiral arm like structures reveal by the JWST can be perceived asymmetric or perturbed in many cases and one SMG (850.030) appears to be undergoing a merger. Indeed, based on the photometric redshifts reported in the CANDELS catalogs \citep{Santini:2015aa,Stefanon:2017aa}, six out of the seven sample SMGs have at least one nearby source with consistent redshifts as those of the corresponding SMGs at an angular distance of 2$''$--5$''$ so $\sim15-45$\,kpc. These evidence suggest that the sample SMGs are experiencing dynamical interactions with nearby sources, which may be responsible in the triggering of their star formation. Similar studies with JWST on a much larger SMG sample will allow determinations of the relative fractions of various morphological classifications such as major merger, minor merger, and isolated disk with or without bars, and these results will provide power constraints to theoretical models.


On the other hand, we measure angular sizes of our sample SMGs in \autoref{sec:cog}. In \autoref{fig:fig4} we show a compilation of all results which are converted into angular physical sizes based on their redshifts. We find median sizes of 7.0$\pm$1.6\,kpc, 6.5$\pm$1.0\,kpc, 5.6$\pm$0.7\,kpc, 5.1$\pm$0.6\,kpc, 4.6$\pm$0.4\,kpc, and 4.3$\pm$0.4\,kpc for F115W, F150W, F200W, F277W, F356W, and F444W, respectively. The reported size measurements of our sample SMGs based on the {\it HST} F125W and F160W \citep{van-der-Wel:2014aa} are also shown, which are consistent with the JWST measurements at similar wavelengths. The sizes of our sample SMGs at F150W are however larger than the reported {\it HST} $H$-band sizes for general SMG samples \citep{Chen:2015aa,Lang:2019aa}, by a factor of about 30\%. This suggests that our sample may be somehow biased large in sizes. This suggestion may help explain a similar amount of difference on the reported F444W sizes between our measurements and those of \citet{Cheng:2022aa}, who found F444W sizes of about 3\,kpc based on two ALMA identified SMGs in the SMACS J0723 field.

The sizes based on NIRCam imaging appear to be significantly larger than those reported based on ALMA observations at 870\,$\mu$m in one of our sample SMG as well as on other samples of SMGs \citep{Simpson:2015aa,Hodge:2016aa,Lang:2019aa,Gullberg:2019aa,Tadaki:2020ab}, which are mostly below 2\,kpc. If we account for the potential size bias of our sample the median F444W size would have been $3.0\pm0.3$\,kpc, still not smaller than the submillimeter sizes and consistent with the stellar sizes of SMGs reported by \citet{Lang:2019aa}, who used $J-H$ colors to infer stellar mass distributions. Our results suggest that while stellar bulges are in active formation phase in SMGs, the total stellar masses are yet dominated by bulges thus the stellar mass sizes, in our case inferred from F444W sizes, are still larger than the star-forming bulges. Our results are in contrast with recent predictions from hydrodynamical simulations, which suggest smaller stellar mass sizes than submillimeter sizes in the mass range of our sample SMGs \citep{Cochrane:2019aa,Popping:2022aa}.

One caveat of our results is that six of the seven SMGs are identified indirectly via radio or MIPS observations, which based on past studies have about 80\% success rate \citep{Hodge:2013lr,Chen:2016aa}. Therefore while a few SMGs are expected to be wrongly identified, we do not expect the results of the overall average sizes to be affected significantly.

Our results demonstrate the power of JWST in understanding the stellar distributions of SMGs. More in-depth analyses on JWST imaging on larger SMG samples should allow us to start addressing issues such as the correlations between sizes and various physical parameters and the properties of clumps and spiral arms. Combining morphological studies using JWST and ALMA would unveil fresh details with regards to how exactly the buildup of stellar bulges in massive galaxies took place at cosmic noon.

\vspace{5mm}
\facilities{JWST(NIRCam), HST(WFC3), JCMT(SCUBA-2), Spitzer(IRAC, MIPS), ALMA}


\software{astropy \citep{Astropy-Collaboration:2018aa},  
          SExtractor \citep{Bertin:1996zr}
          }

\begin{acknowledgements}
C.C.C. acknowledges support from the National Science and Technology Council of Taiwan (NSTC 109-2112-M-001-016-MY3 and 111-2112-M-001-045-MY3). This paper makes use of the following ALMA data: ADS/ JAO.ALMA\#2015.1.01528.S and 2017.1.01027.S. ALMA is a partnership of ESO (representing its member states), NSF (USA) and NINS (Japan), together with NRC (Canada), MOST and ASIAA (Taiwan), and KASI (Republic of Korea), in cooperation with the Republic of Chile. The Joint ALMA Observatory is operated by ESO, AUI/NRAO and NAOJ. This paper was written in a time when the world appears to be in a period of turmoil. We wish peace and stability would soon be upon us all.
\end{acknowledgements}


\bibliography{bib}{}
\bibliographystyle{aasjournal}

\appendix
In the appendix we plot the full {\sc galfit} results in F444W filter in \autoref{fig:figA1} and provide sizes and Sersic indices in \autoref{tab:tableA1}. 
\renewcommand{\thefigure}{A\arabic{figure}}
\setcounter{figure}{0}
\renewcommand{\thetable}{A\arabic{table}}
\setcounter{table}{0}

\begin{figure*}[ht!]
	\begin{center}
		\includegraphics[scale=0.32]{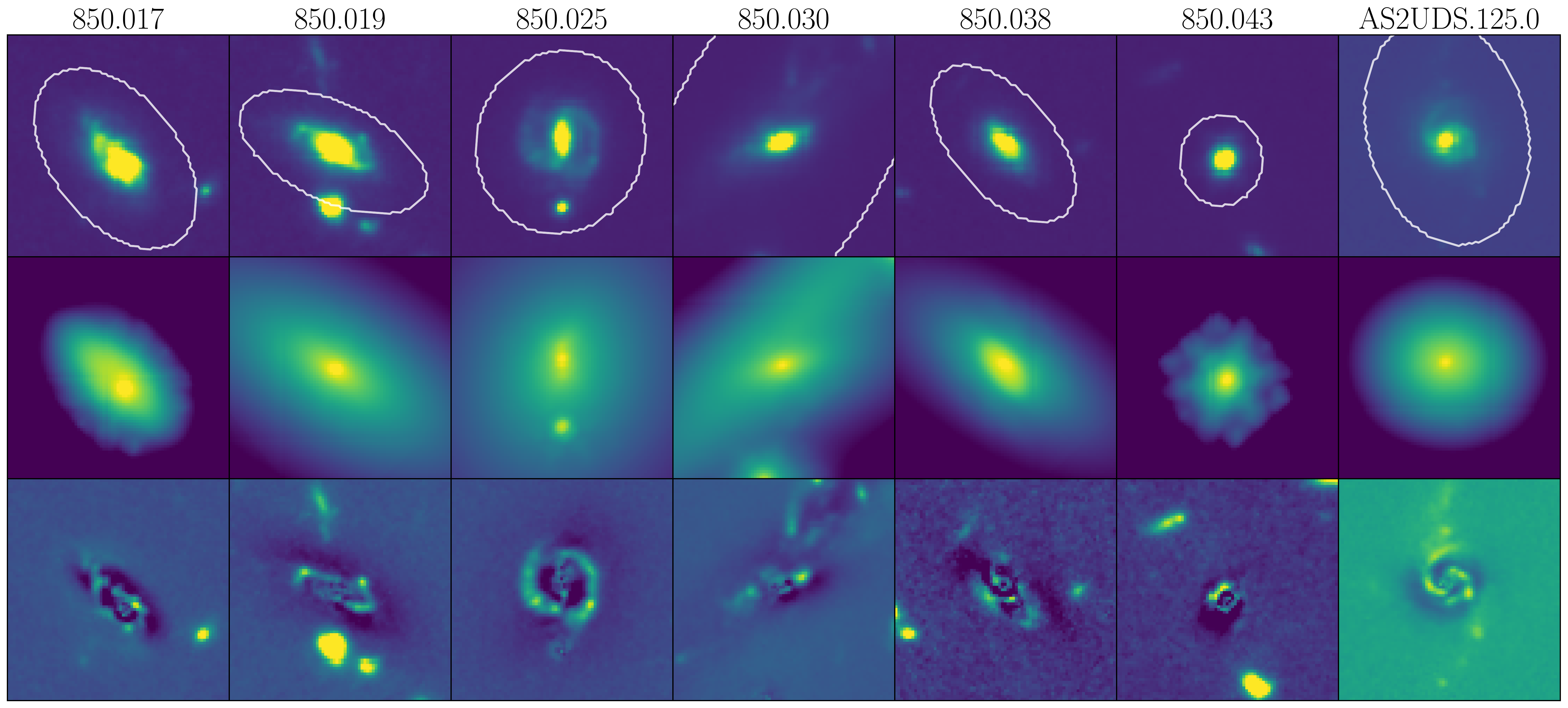}
		\caption{Similar to \autoref{fig:fig2}, but here we show the {\sc galfit} results only in F444W filter but on all sample SMGs.
		}
		\label{fig:figA1}
	\end{center}
\end{figure*}

\begin{table}
    \caption{Results of Sersic fittings using the F444W images}
    \label{tab:tableA1}   
    \begin{center}
        \begin{tabular}{rllll}
            \hline
            ID & \multicolumn{2}{c}{Disk} & \multicolumn{2}{c}{Bulge} \\
             & $R_{\rm e}$ [kpc] & $n$ & $R_{\rm e}$ [kpc] & $n$\\
            \hline
            850.017 & 5.43$\pm$0.05 & 0.44$\pm$0.01 & 0.68$\pm$0.02 & 2.32$\pm$0.10\\
            850.019 & 7.33$\pm$0.60 & 2.14$\pm$0.28 & 0.81$\pm$0.04 & 0.69$\pm$0.11\\
            850.025 & 10.06$\pm$0.56 & 2.54$\pm$0.07 & 0.70$\pm$0.01 & 0.61$\pm$0.04\\
            850.030 & 23.91$\pm$0.60 & 0.61$\pm$0.02 & 3.91$\pm$0.03 & 2.73$\pm$0.02\\
            850.038 & 3.37$\pm$0.02 & 0.33$\pm$0.01 & 4.50$\pm$0.21 & 3.82$\pm$0.10\\
            850.043 & 1.49$\pm$0.22 & 1.33$\pm$0.03 & 0.64$\pm$0.01 & 0.05$^a$\\
            AS2UDS.125.0 & 4.31$\pm$0.03 & 0.79$\pm$0.01 & 0.62$\pm$0.01 & 0.5$^a$\\
            \hline
        \end{tabular}
        \begin{tabular}{l}
            $^a$ Sersic indices are fixed in the fitting to ensure converged results.
        \end{tabular}
    \end{center}
\end{table}

\end{document}